\documentclass[journal,twoside,web]{ieeecolor}

\usepackage{lcsys}
\usepackage{cite}
\usepackage{amsmath,amssymb,amsfonts}

\usepackage{subfigure}

\usepackage{graphicx}

\usepackage{cite}
\usepackage{amsmath}
\usepackage{amssymb,amsfonts}
\usepackage{algorithmic}
\usepackage{graphicx}
\usepackage{textcomp}
\usepackage{xcolor}

\newcommand{\R}{\mathbb{R}}

\newtheorem{theorem}{Theorem}[section]
\newtheorem{proposition}[theorem]{Proposition}
\newtheorem{lemma}[theorem]{Lemma}

\newtheorem{assumption}{Assumption}

\newcommand{\sbs}[2]{{#1}_{\textup{#2}}}

\newcommand{\real}{\mathbb{R}}

\newcommand{\longthmtitle}[1]{\mbox{}{\textit{(#1):}}}

\allowdisplaybreaks[3] 
\renewcommand{\baselinestretch}{1}

\DeclareMathAlphabet{\mymathbb}{U}{BOONDOX-ds}{m}{n}

\newcommand{\zero}{\mymathbb{0}}

\def\BibTeX{{\rm B\kern-.05em{\sc i\kern-.025em b}\kern-.08em
    T\kern-.1667em\lower.7ex\hbox{E}\kern-.125emX}}
\title{Online Regulation of Dynamical Systems to Solutions of Constrained Optimization Problems}

\author{Yiting Chen, Liliaokeawawa Cothren, Jorge Cort\'{e}s, and Emiliano Dall'Anese 
\thanks{Y. Chen, and E. Dall'Anese are with the Department of
Electrical, Computer and Energy Engineering, University of Colorado Boulder, CO, USA;  J. Cort\'{e}s is with the Department of Mechanical and Aerospace Engineering, University of California, San Diego, CA, USA.}
\thanks{This work was supported in part by NSF Awards 2044900 and 2044946, and AFOSR Award FA9550-23-1-0740.}
}

\begin{document}

\maketitle

\thispagestyle{empty} % Removes the page number in the first page
\pagestyle{empty}

\begin{abstract}
This paper considers the problem of regulating a dynamical system to equilibria that are defined as  solutions of an input- and state-constrained optimization problem. To solve this regulation task, we design a state feedback controller based on a continuous approximation of the projected gradient flow. We first show that the equilibria of the interconnection between the plant and the proposed controller correspond to critical points of the constrained optimization problem. We then derive sufficient conditions to ensure that, for the closed-loop system, isolated locally optimal solutions of the optimization problem are locally exponentially stable and show that input constraints are satisfied at all times by identifying an appropriate forward-invariant set.
%Simulations illustrate our results.
\end{abstract}
    
\section{Introduction}
This paper considers the problem of steering the state of a dynamical system to equilibria that are implicitly defined as the solution of a constrained, nonlinear optimization problem.  Such a regulation problem is motivated by optimization and control problems in a number of areas, including  power systems~\cite{li2015connecting}, transportation systems~\cite{bianchin2021time}, epidemic control, and robotics~\cite{carnevale2022aggregative}.
To address this regulation problem, prior work~\cite{Jokic2009controller,brunner2012feedback,Hirata,lawrence2018linear,MC-ED-AB:20, hauswirth2020timescale, nonhoff2021online, bianchin2021time,cothren2022online} in ``online feedback optimization'' has considered the design of controllers based on adaptations of first-order optimization methods. The majority of these works consider equilibria that are solutions to either unconstrained optimization problems or problems with constraints on the control inputs. However, an open research question remains regarding how to systematically design feedback controllers to regulate a dynamic plant to solutions of optimization problems with nonlinear constraints on the system's state, which is what we tackle here.
State constraints are critical across multiple domains: e.g., in power transmission systems, to impose frequency and line flow limits~\cite{li2015connecting}.

\emph{Literature review}. Constrained problems in online regulation of dynamical systems were first considered in~\cite{brunner2012feedback}, where controllers for control-affine systems were engineered based on saddle flows and conditions for asymptotic stability of saddle points of the Lagrangian function were established. Similar gradient-based strategies for constrained convex problems were proposed in~\cite{Hirata}, and local stability results were provided based on non-singularity of the matrix modeling the closed loop.  The Moreau envelope was used in~\cite{MC-ED-AB:20} to deal with linear inequality constraints on the state; however, the stability analysis hinges on augmented Lagrangian approaches, which leads to perturbations of the set of optimal solutions. A primal-dual flow based on a regularized Lagrangian was utilized in~\cite{bianchin2021time}; however, the approach in~\cite{bianchin2021time} is applicable to only linear inequality constraints on the system's state. 
%The work~\cite{hauswirth2018time} considered projected gradient methods for systems modeled as algebraic maps. 
Polytopic constraint sets for the state of linear systems (without disturbances) were considered in for discrete-time linear systems~\cite{nonhoff2021online}.
Finally, our controller design relies on  a continuous approximation of the projected gradient flow, termed safe gradient flow~\cite{allibhoy2023control}, that solves constrained optimization problems. 
The treatment here significantly expands~\cite{allibhoy2023control} by (i)~having the safe gradient flow act as a feedback controller,
(ii) analyzing the stability of the interconnection with a dynamic plant, and (iii) providing a precise characterization of the local stability region.

\emph{Contributions}. We present a new class of feedback controllers for dynamic plants (possibly subject to unknown disturbances) to regulate their input and state to locally optimal solutions of an optimization problem with constraints on the state at steady-state. Our main contribution is twofold. 

\noindent \emph{(a)} We propose a new control design strategy that leverages the safe gradient flow~\cite{allibhoy2023control}. The controller utilizes state feedback to perform the regulation task at hand. 
Critically, our dynamic controller is defined by a locally Lipschitz vector field, thus ensuring
existence and uniqueness of classical solutions, and guarantees that input constraints are enforced at all times; moreover, its equilibria correspond to the critical points of the optimization problem. 

\noindent \emph{(b)} We show the existence of a forward invariant set for the interconnection of the dynamic plant and the proposed controller, and we leverage singular perturbation theory~\cite{khalil2002nonlinear}  to find sufficient conditions for the stability of the closed-loop system. We show that isolated locally optimal solutions of the optimization problem are locally exponentially stable and characterize the region of attraction.

\section{Preliminaries and Problem Formulation}

\emph{Notation}.  $(\cdot)^\top$ denotes transposition. For a given vector $x\in\mathbb{R}^n$, $\|x\| :=\sqrt{x^\top x}$. Given the  vectors $x\in\mathbb{R}^n$ and $y\in\mathbb{R}^m$, $(x,y)\in\mathbb{R}^{n + m}$ denotes their concatenation. For a smooth function $f:\mathbb{R}^n\rightarrow \mathbb{R}$, the gradient is denoted by $\nabla f$ and its Hessian matrix is denoted by $\nabla^2 f$. For a continuously differentiable function $h:\mathbb{R}^n\rightarrow \mathbb{R}^m$, its Jacobian matrix is denoted by $J_h$. For any natural number $n$, $[n]$ denotes the set $\{1,\cdots,n \}$. For any vectors, $a$ and $b$, $a\leq b$ means that all the entries of $a-b$ are less than or equal to $0$. The distance between a point $s$ and an nonempty set $\mathcal{S}$ is defined as $\operatorname{dist}(s,\mathcal{S})=\inf_{s_0\in\mathcal{S}} \|s-s_0\|$. The diameter of a set $\mathcal{S}$ is defined as $\operatorname{diam}(\mathcal{S}):=\sup_{s_1,s_2\in\mathcal{S}} \|s_1-s_2\| $.  We define $\mathcal{B}_n(x^*, r) := \{x \in \R^{n} : \|x - x^*\| < r\}$ and $\zero_n \in \R^n$ the vector of all zeros.

\emph{Plant model}. We consider  systems that can be modeled using continuous-time dynamics
\begin{align}
\label{eq:plantModel}
 \dot x & =  f(x, u, w), \hspace{.4cm} x(t_0) = x_0 
\end{align}
where 
$f: \mathcal{X} \times \mathcal{U}\times \mathcal{W} \rightarrow \real^{n}$, with
$\mathcal{X}\subseteq \real^{n}$, $\mathcal{U} \subseteq \real^{n_u}$, and $\mathcal{W} \subseteq \real^{n_w}$ open and connected sets. 
In~\eqref{eq:plantModel}, ${x\in\mathcal{X}}$ is the state (with 
$x_0 \in \mathcal{X}$ the initial condition),  
$u\in \mathcal{U}_c \subset \mathcal{U}$  is the control input, and $w\in \mathcal{W}_c \subset \mathcal{W}$ is an unknown disturbance. We assume that $\mathcal{U}_c$ and $\mathcal{W}_c$ are compact, and that the vector 
field $f$ is continuously differentiable and Lipschitz-continuous. We make the following assumptions. %\footnote{\emph{Notation}.  $(\cdot)^\top$ denotes transposition. For a given vector $x\in\mathbb{R}^n$, $\|x\| :=\sqrt{x^\top x}$. Given the  vectors $x\in\mathbb{R}^n$ and $y\in\mathbb{R}^m$, $(x,y)\in\mathbb{R}^{n + m}$ denotes their concatenation. For a smooth function $f:\mathbb{R}^n\rightarrow \mathbb{R}$, the gradient is denoted by $\nabla f$ and its Hessian matrix is denoted by $\nabla^2 f$. For a continuously differentiable function $h:\mathbb{R}^n\rightarrow \mathbb{R}^m$, its Jacobian matrix is denoted by $J_h$. For any natural number $n$, $[n]$ denotes the set $\{1,\cdots,n \}$. For any vectors, $a$ and $b$, $a\leq b$ means that all the entries of $a-b$ are less than or equal to $0$. The distance between a point $s$ and an nonempty set $\mathcal{S}$ is defined as $\operatorname{dist}(s,\mathcal{S})=\inf_{s_0\in\mathcal{S}} \|s-s_0\|$. The diameter of a set $\mathcal{S}$ is defined as $\operatorname{diam}(\mathcal{S}):=\sup_{s_1,s_2\in\mathcal{S}} \|s_1-s_2\| $.  We define $\mathcal{B}_n(x^*, r) := \{x \in \R^{n} : \|x - x^*\| < r\}$ and denote by $\zero_n \in \R^n$ the vector of all zeros.} 

\begin{assumption}[\textit{Steady-state map}]
\label{as:steadyStateMap}
There exists a unique continuously differentiable function $h:\mathcal{U}\times\mathcal{W}  \to \mathcal{X}$ 
such that, for any fixed $\bar u \in \mathcal{U}$ and $\bar w \in \mathcal{W}$, 
$f(h(\bar u, \bar w), \bar u, \bar w) = 0$.
Moreover, $h(u,w)$ admits the decomposition $h(u,w)=h_u(u)+h_w(w)$, and  $h_w(w)$ and the Jacobian $J_h(u):=\frac{\partial h_u(u)}{\partial u}$ are locally Lipschitz continuous.  \hfill $\Box$
\end{assumption}

Assumption~\ref{as:steadyStateMap} guarantees that, due to the continuity of $J_h(u)$  and compactness of $\mathcal{U}_c$ and $\mathcal{W}_c$, there exists $\ell_{h_u}, \ell_{h_w}\geq 0$ such that $\left\|J_h(u)\right\|\leq\ell_{h_u}$ and $\|h_w(w_1)-h_w(w_2)\|\leq \ell_{h_w}\|w_1-w_2\|$  hold for any $u\in\mathcal{U}_c $ and any $w_1,w_2\in\mathcal{W}_c$. 
Hereafter, we denote the compact set of admissible equilibrium points of the system~\eqref{eq:plantModel} by $\sbs{\mathcal{X}}{eq} := h(\mathcal{U}_c\times\mathcal{W}_c)$. 
Let $r > 0$ denote the largest positive constant such that
$\mathcal{X}_r :=\sbs{\mathcal{X}}{eq} + \mathcal{B}_n(\zero_n,r)$ satisfies  $\mathcal{X}_r \subseteq  \mathcal{X}$. We make the following stability assumption on the plant, which is common in the feedback optimization literature~\cite{hauswirth2020timescale,MC-ED-AB:20,lawrence2018linear}.

\begin{assumption}[\textit{Stability}]
\label{as:stabilityPlant}
$\exists ~ a, k >0$ such that, for any fixed 
$\bar u \in \mathcal{U}_c$  and $\bar w\in \mathcal{W}_c$, the bound
%\begin{align}   
%\label{eq:expstability}
$\|x(t) - h(\bar u, \bar w)\| \leq 
k \|x_0 - h(\bar u, \bar w)\| e^{-a (t - t_0)}$, 
%\end{align}
holds for all $t \geq t_0$, for some $t_0 \geq 0$, and for every initial condition 
$x_0 \in \mathcal{X}_0 := \sbs{\mathcal{X}}{eq} + \mathcal{B}_n(\zero_n, r_0)$, $r_0<r/k-\operatorname{diam}(\mathcal{X}_{eq})$, 
where $x(t)$ is the solution of \eqref{eq:plantModel}. 
\hfill $\Box$
\end{assumption}

Assumption \ref{as:stabilityPlant} guarantees the equilibrium $h( u,w)$ is exponentially
stable, uniformly in time. Using this, the existence of a Lyapunov function is guaranteed by the following result, which is a slight extension of~\cite[Prop. 2.1]{hauswirth2021corrigendum}.

\begin{lemma}\longthmtitle{Existence of a Lyapunov function}
\label{lem:converse_plant}
Consider the system~\eqref{eq:plantModel} satisfying 
Assumptions~\ref{as:steadyStateMap}-\ref{as:stabilityPlant}, where $\mathcal{X}_0$ is the set of initial conditions defined in Assumption~\ref{as:stabilityPlant}.
Then, for any fixed $w \in \mathcal{W}_c$, there exists a function  
$W: \mathcal{X}_0 \times \mathcal{U} \to \R$ that satisfies 
the inequalities: 
\begin{align}
& d_1 \|x - h(u,w)\|^2  \leq W(x,u) \leq d_2 \|x - h(u,w)\|^2, \nonumber\\
& \frac{\partial W}{\partial x}f(x,u,w)  \leq -d_3\|x - h(u,w)\|^2, \nonumber  \\ 
&\left\| \frac{\partial W}{\partial x}\right\| \leq d_4 \|x - h(u,w)\|, ~~ \left\| \frac{\partial W}{\partial u}\right\| \leq d_5 \|x - h(u,w)\|,\nonumber
\end{align}
for some positive constants $d_1\leq d_2, d_3, d_4,d_5$.  \hfill $\triangle$
\end{lemma}

Our control problem is formalized next.  

\emph{Target control problem}. We consider an optimization problem of the form: 
\begin{subequations}
\label{eq:optimization-problem-2}
\begin{align}
        \min_{ u\in\mathbb{R}^{n_u}}\quad &\phi(u) + \psi\left( h(u,w) \right)\\
     \text{s.t. } & \ell \left( h(u,w) \right) \leq 0, ~~ \gamma(u)\leq 0 
\end{align}
\end{subequations}
where the functions $\phi: \mathbb{R}^{n_u} \rightarrow \mathbb{R}$,  $\psi : \mathbb{R}^{n} \rightarrow \mathbb{R}$, and $\ell: \mathbb{R}^{n} \rightarrow \mathbb{R}^p$ have a locally Lipschitz continuous gradient (Jacobian), and where $\gamma(u)=[\gamma_1(u),\cdots,\gamma_m(u)]^\top : \mathbb{R}^{n_u} \rightarrow \mathbb{R}^m$ is continuously-differentiable. 
We assume that $\mathcal{U}_c$ can be expressed as $\mathcal{U}_c= \{u: \gamma(u) \leq 0\}$. We note that the constraints $\ell(x) \leq 0$ specify a given desirable set for the state of the system at steady state; we also notice that this set is parametrized by the unknown disturbance $w$ since it can be rewritten as $\ell(h(u,w)) \leq 0$. The presence of this constraint is a key differentiating factor relative to existing works on online feedback optimization~\cite{brunner2012feedback,lawrence2018linear,MC-ED-AB:20, hauswirth2020timescale, bianchin2021time}.

We make the following assumptions on~\eqref{eq:optimization-problem-2}.

\begin{assumption}[\textit{Set of inputs}]
\label{as:compactnessofUc}
     For any $i \in [m]$ and any $u\in\mathcal{U}_c$, it holds that $\nabla \gamma_i(u)\neq 0$ if $\gamma_i(u)=0$.\hfill $\Box$
\end{assumption}

\begin{assumption}[\textit{Regularity}]
\label{as:openloop}
    Let $u^* \in \mathcal{U}_c$ be a local minimizer and an isolated 
    %KKT
    Karush–Kuhn–Tucker (KKT)
    point for the optimization problem~\eqref{eq:optimization-problem-2}. The following hold: 

    \noindent \textit{i)}
    Strict complementarity condition~\cite{fiacco1976sensitivity} and the linear independence constraint qualification (LICQ) hold at $u^*$.
    
    \noindent \textit{ii)}
    The maps $u \mapsto \gamma(u)$, $u \mapsto \phi(u)$, $u \mapsto \psi(h(u,w))$, and $u \mapsto \ell(h(u,w))$ are twice continuously differentiable over some open neighborhood of $u^*$ and their Hessian matrices are positive semi-definite at $u^*$.

    \noindent \textit{iii)}
    The Hessian $\nabla^2 \phi(u^*)$ is positive definite.
     \hfill $\Box$
\end{assumption}

Assumption~\ref{as:compactnessofUc}  is satisfied when $\mathcal{U}_c = \{u: \|u-u_0\|_p \leq r\}$ for a given $r > 0$ and for $1\leq p\leq +\infty$, or when $\mathcal{U}_c$ is a polytope; it is also satisfied in applications such as the ones described in~\cite{li2015connecting,MC-ED-AB:20,carnevale2022aggregative,bianchin2021time}. Assumption~\ref{as:openloop} is satisfied when~\eqref{eq:optimization-problem-2} is convex with a strongly convex function~\cite{MC-ED-AB:20,carnevale2022aggregative}; here, we provide a minimal set of assumptions that allows us to consider non-convex problems while still allowing for strong stability guarantees as discussed in the next section. We refer the reader to~\cite{Beck:2014} for the notions of local minimizer and KKT point. Next, we outline our problem. 

\textit{Problem (Regulation to optimal solutions}): Design a feedback controller to regulate inputs and states 
of~\eqref{eq:plantModel} to a minimizer $u^*$ of~\eqref{eq:optimization-problem-2} and the optimal state $x^* = h(u^*,w)$ without requiring knowledge of the disturbance $w$, while respecting input constraints at all times. 
 \hfill $\Box$

\section{Controller Design and Stability Analysis}

%We outline our proposed state-feedback controller to solve our regulation problem and characterize the stability of the closed-loop system.

\subsection{Approximate projected gradient controller}
To solve our regulation problem, 
we propose the following  state-feedback controller: 
\begin{align}
\label{eq:controller}
         &\dot u =\eta F_\beta(x,u)\\
\begin{split}\label{eq: F(x,u)}
        &  F_\beta(x,u): = \arg\min_{\theta} \|\theta + \nabla \phi(u) + J_h^\top \nabla \psi(x)\|_2^2, \\
        & ~~~~~~~~~~~~\text{s.t. } \frac{\partial \ell}{\partial x}(x) J_h(u) \theta \leq -\beta \ell(x),
        \\ & ~~~~~~~~~~~~~~~~\frac{\partial \gamma}{\partial u}(u) \theta \leq - \beta \gamma(u),
    \end{split}
\end{align}
where $\beta \in \R_{>0}$ is a design parameter and $\eta > 0$ is the controller gain. 
To gain intuition on this design, we note that~\eqref{eq:controller} is an approximation of the projected gradient flow 
$\dot u = \text{proj}_{T_\mathcal{F}(u)}(- \nabla \phi(u) - J_h^\top \nabla \psi(h(u,w)))$, where $T_\mathcal{F}(u)$ denotes the tangent cone of $\mathcal{F}(u) := \{u: \ell \left( h(u,w) \right) \leq 0, ~~ \gamma(u)\leq 0\}$ at $u$; 
in fact, one can show~\cite[Prop. 4.4]{allibhoy2023control} that $\lim_{\beta \rightarrow \infty} F_\beta(h(u,w),u) = \text{proj}_{T_\mathcal{F}(u)}(- \nabla \phi(u) - J_h^\top \nabla \psi(h(u,w)))$. A key modification relative to the projected gradient flow is that the steady-state map $h(u,w)$ is replaced by measurements of the system state $x$; this allows us to leverage measurements of the state to steer it to the solution $(u^*, x^*)$ of the problem~\eqref{eq:optimization-problem-2} without requiring knowledge of $w$. 

\begin{assumption}[\textit{Feasibility and conditions}]
\label{as:feasibility}
    For all $x\in \mathcal{X}$ and $ u \in \mathcal{U}_c$, $\exists~ \theta \in \mathbb{R}^{n_u}$ such that $\frac{\partial \ell}{\partial x}(x)J_h(u) \theta \leq -\beta \ell(x) \text{ and } \frac{\partial \gamma}{\partial u}(u) \theta \leq - \beta \gamma(u)$. For any $x\in \mathcal{X}$ and $ u \in \mathcal{U}_c$,~\eqref{eq: F(x,u)} satisfies the Mangasarian-Fromovitz Constraint Qualification and the constant-rank condition~\cite{liu1995sensitivity}.\hfill $\Box$
\end{assumption}

Since the constraints in~\eqref{eq: F(x,u)} are based on techniques from Control Barrier Functions (CBFs)~\cite{allibhoy2023control}, Assumption~\ref{as:feasibility} guarantees that there always exists a direction that keeps the system inside the feasible set of problem~\eqref{eq:optimization-problem-2}. 
We show later that Assumption~\ref{as:feasibility} can  be weakened to a subset of $\mathcal{X}$. 
Defining $z := (x,u)$, the plant~\eqref{eq:plantModel} under the controller~\eqref{eq:controller} leads to the following interconnected system: 
\begin{align}
    \label{eq:interconnected-system}
    \dot{z} = F(z,w), ~~~~~ F(z,w) := \begin{bmatrix} f(x,u,w)
    \\
    \eta F_\beta(x,u)
    \end{bmatrix}, 
\end{align}
with initial condition $z(t_0) = (x(t_0), u(t_0))$. Before presenting our main convergence and stability results for~\eqref{eq:interconnected-system}, we discuss some important properties. The proof of all results is postponed to Section~\ref{sec:proofs}.

\vspace{.1cm}

\begin{proposition}[\textit{Forward invariance}]
\label{forward-invariant-Uc-proposition}
 Let Assumptions \ref{as:compactnessofUc} and \ref{as:feasibility} be satisfied.
 Then, \eqref{eq:interconnected-system} renders the set $\mathcal{U}_c$ forward invariant.  \hfill $\triangle$
\end{proposition}
%
%\marginJC{It's actually more than this, no? Isn't it the set $\mathcal{F}(u)$ which is forward invariant? Also, I haven't read the intro yet, but it seems like we should make emphasis that this is precisely what the sgf gives you: anytime satisfaction of constraints on inputs. \blue{Reply: No, the set $\mathcal{F}(u)$ is not forward invariant for \eqref{eq:interconnected-system}. $\mathcal{F}(u)$ is forward invariant for $\dot u = \text{proj}_{T_\mathcal{F}(u)}(- \nabla \phi(u) - J_h^\top \nabla \psi(h(u,w)))$. Note that we replace $h(u,w)$ with $x$ in  \eqref{eq:interconnected-system}, resulting in the loss of forward- invariantness.}}
%

\vspace{.1cm}

\begin{proposition}[\textit{Lipschitzness}]   
\label{LipschitzofF}
    Let Assumptions~\ref{as:openloop}-\ref{as:feasibility} be satisfied. Then:
    
    \noindent (i) for any $w \in \mathcal{W}$, $u \mapsto F_\beta(h(u,w),u)$ is Lipschitz continuous with constant $\ell_{F_u} \geq 0$ over $\mathcal{U}_c$;
    
    \noindent (ii) for any $u\in\mathcal{U}_c$, $x \mapsto F_\beta(x,u)$ is locally Lipschitz continuous;
    
    \noindent (iii) For any compact subset $\tilde{\mathcal{X}}\subseteq \mathcal{X}$ and any $x\in \tilde {\mathcal{X}}$, $u \mapsto F_\beta(x,u)$ is locally Lipschitz continuous.   \hfill $\triangle$
\end{proposition}

\vspace{.1cm}

Proposition~\ref{forward-invariant-Uc-proposition} guarantees that constraints on the inputs are satisfied, while Proposition~\ref{LipschitzofF} ensures the existence and uniqueness  solutions in the classical sense; this is a key advantage over projected gradient flows which may be, in general, discontinuous.

\subsection{Stability analysis}\label{sec:analysis-for-interconnnected-system}

This section characterizes the  stability of~\eqref{eq:interconnected-system}. 
We first establish the existence of a compact and forward-invariant set for  the state~$x$; this is necessary for the Lipschitz constant in Proposition~\ref{LipschitzofF}(iii) to be well defined and plays an integral part in the proof of the main stability result. 

Define the compact set $\mathcal{X}_1:=\{x\mid \operatorname{dist}(x,\mathcal{X}_{eq}) < \sqrt{d_2/d_1}\left(d_0+\operatorname{diam}(\mathcal{X}_{eq}) \right)\}$, where
   \begin{align*}
       d_0 :=\max \left\{\operatorname{dist}(x(t_0),\mathcal{X}_{eq}), \frac{2 (d_4\ell_{h_u}+d_5) \ell_{F_u}  M_u}{d_3}\alpha_0 \right\}, 
   \end{align*}
with $t_0\geq 0$, $\alpha_0>0$ and $M_u :=\max_{u\in \mathcal{U}_c} \|u-u^*\| $. Proposition~\ref{LipschitzofF}(iii) applies to $\mathcal{X}_1$ with $t_0=0$;
we denote as $\ell_{F_x}$ the Lipschtiz constant of $F_\beta(x,u)$ w.r.t $x$ over $\mathcal{X}_1$. The following result establishes forward invariance of $\mathcal{X}_1$.

\vspace{.1cm} 

\begin{lemma}[Forward invariance]   
\label{forward-invariant-X}
Consider system~\eqref{eq:interconnected-system} and let Assumptions \ref{as:steadyStateMap}, \ref{as:stabilityPlant}, and \ref{as:feasibility} hold. Assume that $r_0>\sqrt{d_2/d_1}\operatorname{diam}(\mathcal{X}_{eq})$. If $\operatorname{dist}(x(t_0),\mathcal{X}_{eq})$$ \leq \sqrt{d_1/d_2} r_0-\operatorname{diam}(\mathcal{X}_{eq})$ and $$\alpha_0\leq   \frac{d_3}{2 ( \ell_{F_x}\ell_{h_u}+d_5) \ell_{F_u}  M_u}\left(\sqrt{d_1/d_2} r_0-\operatorname{diam}(\mathcal{X}_{eq}) \right).$$ Then: 

\noindent (a) $\mathcal{X}_1\subseteq \mathcal{X}_0$; and, 

\noindent  (b) for any  $\eta\leq\min \{ \frac{d_3}{2 (d_4\ell_{h_u}+d_5) \ell_{F_x} },\alpha_0 \}$, the state $x(t)$ never leaves $\mathcal{X}_1$ after time $t\geq t_0$. 
\hfill $\triangle$
\end{lemma}

\vspace{.1cm} 

Note that, since $\mathcal{X}_1$  is forward invariant,  Assumption~\ref{as:feasibility} can be restricted to $\mathcal{X}_1$. Additionally, by comparing the 
%Karush–Kuhn–Tucker (KKT)  
KKT
conditions for \eqref{eq:optimization-problem-2} and for the optimization defining $F_\beta$, we obtain the following result. 

\vspace{.1cm} 

\begin{proposition}[Equilibria and optimizer]   
\label{prop:equivalence}
    There exists $\lambda^*$ such that $(u^*,\lambda^*)$ is a KKT point for \eqref{eq:optimization-problem-2}  if and only if $(h(u^*,w),u^*)$ is an  equilibrium for \eqref{eq:interconnected-system}.
    \hfill $\triangle$
\end{proposition}

\vspace{.1cm} 

Before stating the main stability result, we introduce some useful notation. Let $\tilde z=(x-x^*,u-u^*)$  and define 
$E:=\frac{\partial F_\beta (h(u,w),u)}{\partial u}\mid_{u=u^*}$, $
e_1 :=-\lambda_{\max}(E)$, and $e_2 :=-\lambda_{\min}(E)$. Then, we can write the dynamics as~\cite{khalil2002nonlinear}: $F_\beta( h(u,w),   u) = E (u-u^*) + \hat{g}(u)$, where $\hat{g}(u)$ satisfies $\|\hat{g}(u)\|_2\leq L \|u-u^*\|_2^2$, $\forall u\in \mathcal{B}_{n_u}(u^*,\delta)$, for some $L> 0$ and $\delta>0$. Define
$$s_{\min}=\left\{\begin{array}{lll} 0&, & \text{ if } \delta\geq \frac{e_1}{L}, \\ 
1-\frac{\delta L}{e_1}&, & \text{ if } \delta<\frac{e_1}{L} .
\end{array}\right.$$
Also, let  $M \in \mathbb{R}^{2 \times 2}$ with entries  $m_{11}=  \frac{\theta}{\eta}(d_3-d_4\ell_{h_u} \ell_{F_x}\eta-d_5 \ell_{F_x}\eta)$, $m_{12}=m_{21}= -\frac{1}{2} ( \theta (d_4\ell_{h_u}+d_5) \ell_{F_u}+(1-\theta) \frac{\kappa \ell_{F_x}}{e_1} )$, $m_{22}=  (1-\theta)\kappa s$, where  $\theta=\kappa \ell_{F_x}(e_1\ell_{F_u}(d_4\ell_{h_u}+d_5)+\kappa \ell_{F_x})^{-1} $ and $\kappa > 0$.

\vspace{.2cm}

\begin{theorem}[\textit{Local exponential stability}] 
\label{thm:stability}
    Consider the system~\eqref{eq:interconnected-system} satisfying  Assumption \ref{as:steadyStateMap}-\ref{as:feasibility}, and let $(x(t),u(t))$, $t \geq t_0$,  be the unique trajectory of~\eqref{eq:interconnected-system}. Assume that $r_0>\sqrt{d_2/d_1}\operatorname{diam}(\mathcal{X}_{eq})$ and let $\alpha_0$ satisfy the conditions on Lemma~\ref{forward-invariant-X}.    Then, for any $\kappa>0$,  any  $s\in(s_{\min},1]$, 
    and $0< \eta < \min\left\{\eta^*_1, \eta^*_2, \alpha_0 \right\}$, with 
    $$\eta^*_1 := \frac{sd_3e_1}{ \ell_{F_x}(d_4\ell_{h_u}+d_5)(\ell_{F_u}+e_1s)},  \eta^*_2 :=  \frac{d_3}{2 (d_4\ell_{h_u}+d_5) \ell_{F_x} } $$ 
    it holds that $M$ is positive definite and
    \begin{align}  
\left\|\tilde z(t)\right\|  
 \leq \bar{r} \left\|\tilde z(t_0)\right\|  e^{-\frac{1}{2}\lambda_M r_2(t-t_0)}, ~\forall~  t \geq t_0,
 \end{align}
 where $\bar r := \sqrt{\frac{r_1}{r_2}}  (1+\ell_{h_u}^2+\ell_{h_u})$, $r_1, r_2$ are defined as
\begin{align*}
    r_1 := \max \left\{\frac{\eta}{\theta d_1},\frac{2e_2\eta}{\kappa(1-\theta)} \right\} , r_2 := \min \left\{ \frac{\eta}{\theta d_2},\frac{2e_1\eta}{\kappa(1-\theta)} \right\}
\end{align*}
and $\lambda_M = \lambda_{\min}(M)$, for any initial condition $(x(t_0),u(t_0))$ such that $\operatorname{dist}(x(t_0),\mathcal{X}_{eq})$$ \leq \sqrt{\frac{d_1}{d_2}} r_0-\operatorname{diam}(\mathcal{X}_{eq})$ , $\|u(t_0)-u^*\|\leq \frac{e_1}{L}(1-s)$. 
\hfill $\triangle$
\end{theorem}

\vspace{.1cm}

Theorem~\ref{thm:stability} establishes local exponential stability of $(u^*, x^*)$, where we recall that $u^*$ satisfies Assumption~\ref{as:openloop} and $x^* = h(u^*,w)$. We note that the free parameter $s$ affects both $\lambda_M$ and the size of the region of attraction; in particular, as $s$ decreases, the region gets smaller and $\lambda_M$ may increase. We also note that the other free parameter $\kappa$ can be used to maximize $\lambda_M r_2$. However, this is something that may be burdensome for a numerical perspective. The result of Theorem~\ref{thm:stability} holds for constant disturbances; the extension to time-varying disturbances will be the subject of future research. If the QP problem~\eqref{eq: F(x,u)} is not solved to convergence, then we would have an inexact implementation of the controller; in this case, by combining Theorem~\ref{thm:stability} and~\cite[Lemma~9.4]{khalil2002nonlinear}, it is possible to derive results in terms or practical local exponential stability.

\section{Proofs}
\label{sec:proofs}

For brevity, we use the  shorthand notation $F_\beta(u):=F_\beta(h(u,w),u)$. %Moreover, we will use 
Consider the variable shift $\tilde x = x-h(u,w)$, which shifts the equilibrium of \eqref{eq:plantModel} to
the origin. In the new variables, \eqref{eq:interconnected-system} reads as:
\begin{subequations}\label{eq:interconnected-system_change_variable}
\begin{align}
   \label{eq:interconnected-system-a_change_variable}
    \dot{\tilde{x}} &=f(\tilde{x}+h(u,w), u,w)-\frac{d}{d t} h(u,w)\\
    \begin{split}\label{eq:interconnected-system-b_change_variable}
        \dot u &=\eta F_\beta(\tilde x+h(u,w),u)
    \end{split}
\end{align}
\end{subequations}
For \eqref{eq:interconnected-system-a_change_variable}, we denote by $W ( \tilde x, u)$ the Lyapunov
function from Lemma~\ref{lem:converse_plant}.

\vspace{.1cm}

\emph{(a) Proof of Proposition \ref{forward-invariant-Uc-proposition}}. By definition, $\frac{\partial \gamma}{\partial u}F_\beta(x,u)\leq -\beta \gamma(u)$ or, equivalently $\nabla \gamma_i^\top F_\beta(x,u)\leq -\beta \gamma_i(u)$. Then, using~\eqref{eq:controller}, for  $ u \in \partial \mathcal{U}_c=\cup_{i}\{u: \gamma_i(u) = 0\}$, we have $\dot \gamma_i(u)=\nabla \gamma_i^\top F_\beta(x,u)\leq -\beta \gamma_i(u)=0$. Hence by Nagumo’s Theorem~\cite{nagumo1942lage},
$\{u:\gamma_i(u)\leq 0\}$ is forward invariant for all $i$.

\vspace{.1cm}

\emph{(b) Proof of Proposition~\ref{LipschitzofF}}. Note that the statement describes the (local) Lipschitzness of both $F_\beta(x,u)$ and $F_\beta(h(u,w),u)$. To establish these properties, we apply~\cite[Theorem 3.6]{liu1995sensitivity} to show  the local Lipschitzness of functions defined by quadratic programs with parameter-dependent linear inequality constraints. 

For (i), note that~\eqref{eq: F(x,u)} with $x$ replaced by $h(u,w)$ (with $h(u,w) \in \mathcal{X}$) satisfies
the General Strong Second-Order Sufficient Condition~\cite{liu1995sensitivity} and Slater’s condition at $u \in \mathcal{U}$. Moreover, the cost and constraints of \eqref{eq: F(x,u)} are twice continuously differentiable. Thus, $F_\beta(h(u,w),u)$ is locally Lipschitz at $u \in \mathcal{U}$ by~\cite[Theorem 3.6]{liu1995sensitivity}. Finally, $F_\beta(h(u,w),u)$ is Lipschitz on $\mathcal{U}_c$ due to the compactness of $\mathcal{U}_c$. A similar reasoning can be applied to prove the local Lipschitzness of $F_\beta(x,u)$ in each of its arguments.

%\hfill $\Box$ 

\vspace{.1cm}

% ancla
\emph{(c) Proof of Lemma \ref{forward-invariant-X}}. First, it is straightforward to verify that  $\mathcal{X}_1 \subseteq \mathcal{X}_0$. Next, define $\mathcal{X}_2=\{x\mid \inf_{x\prime \in \mathcal{X}_{eq} }\|x-x^\prime\| \leq d_0\}$, then $x(t_0)\in\mathcal{X}_2 \subsetneqq \mathcal{X}_1$. We show that for  $\eta\leq\min \{ \frac{d_3}{2 d_4\ell_{h_u} \ell_{F_x} },\alpha_0 \}$, we have $\frac{d}{d t} W(\tilde x, u)<0$ for any $x(t)\in \mathcal{X}_1\setminus \mathcal{X}_2$. Note that,
\begin{subequations}
\begin{align}
 &\frac{d}{d t} W(\tilde x, u)=\frac{\partial W}{\partial \tilde{x}}\dot{\Tilde{x}}+\frac{\partial W}{\partial u} \dot u 
 \end{align}
 
   \begin{align}
=&  \frac{\partial W}{\partial \tilde{x}} f(\tilde{x}+h(u,w), u)- \frac{\partial W}{\partial \tilde{x}}J_{h}(u)\dot u +\frac{\partial W}{\partial u} \dot u   
\end{align}
 \begin{align}
\leq & -d_3\|\tilde{x}\|^2+\left(d_4 \ell_{h_u}+d_5\right)\|\dot{u}\|\|\tilde{x}\| 
\end{align}
\end{subequations}

Next, we bound $\|\dot u\|$. If  $x\in\mathcal{X}_1$, then
\begin{align*}
    \|\dot u\|&=\eta\|F_\beta(x,u)\| 
    =  \eta\|F_\beta(x,u)-F_\beta(u^*)\|\\
    & \leq  \eta \|F_\beta(x,u)-F_\beta(u)\| +\eta \|F_\beta(u)-F_\beta(u^*)\|\\
    & \leq  \eta  \ell_{F_x} \|x-h(u,w)\| +\eta\|F_\beta(u)-F_\beta(u^*)\|
\end{align*}
and note that $\| F_\beta(u)-F_\beta(u^*)\|\leq \ell_{F_u} \|u-u^*\|$. Hence, if  $x\in\mathcal{X}_1$, one has that 
\begin{align*}
&\frac{d}{d t} W(\tilde x, u)    \\
\leq & -d_3\|\tilde{x}\|^2+\left(d_4 \ell_{h_u}+d_5\right)\|\tilde{x}\|( \ell_{F_x} \eta \|\tilde{x}\|+\ell_{F_u} \eta \|u-u^*\|) \\
\leq& (-d_3+ (d_4\ell_{h_u} \hspace{-.15cm} +d_5) \ell_{F_x} \eta   )\|\tilde{x}\|^2 +   \left(d_4 \ell_{h_u} \hspace{-.15cm} +d_5\right) \ell_{F_u} \eta M_u\|\tilde{x}\|. 
\end{align*}
It then follows that $\frac{d}{d t} W(\tilde x, u) < 0$ if 
\begin{align*}
\|\tilde x\|> \frac{\left(d_4 \ell_{h_u}+d_5\right) \ell_{F_u} \eta M_u}{d_3- (d_4\ell_{h_u}+d_5) \ell_{F_x} \eta} \text{~and~} \eta<\frac{d_3}{(d_4\ell_{h_u}+d_5) \ell_{F_x}}.
\end{align*}
For any $x(t)\in \mathcal{X}_1\setminus \mathcal{X}_2$ and any $\eta\leq\min \{ \frac{d_3}{2 (d_4\ell_{h_u}+d_5) \ell_{F_x} },\alpha_0 \}$, one has that 
\begin{align*}
    \|\tilde x(t)\| & > d_0\geq  \frac{2 (d_4\ell_{h_u}+d_5) \ell_{F_u}  M_u}{d_3}\alpha_0 \\
    & \geq \frac{ (d_4\ell_{h_u}+d_5) \ell_{F_u}  M_u \eta}{\frac{d_3}{2}}\geq \frac{(d_4\ell_{h_u}+d_5) \ell_{F_u} \eta M_u}{d_3- (d_4\ell_{h_u}+d_5) \ell_{F_x} \eta},
\end{align*}
   implying that $\frac{d}{d t} W(\tilde x, u)<0$.
   
Next, we show that $x(t)$ will not exit $\mathcal{X}_1$. Otherwise, there must exist $t_2>0$ such that 
    \[\inf_{x\prime \in \mathcal{X}_{eq} }\|x(t_2)-x^\prime\| = \sqrt{d_2/d_1}\left(d_0+\operatorname{diam}(\mathcal{X}_{eq}) \right).\]
    Additionally, we know that $\inf_{x\prime \in \mathcal{X}_{eq} }\|x(0)-x^\prime\| \leq d_0$,
    by continuity, there exists $t_1$ such that $0<t_1<t_2$, $x(t) \in \mathcal{X}_1\setminus \mathcal{X}_2$,  ~$\inf_{x\prime \in \mathcal{X}_{eq} }\|x(t)-x^\prime\| > d_0+\epsilon, ~\forall t\in [t_1,t_2)$,  and $\inf_{x\prime \in \mathcal{X}_{eq} }\|x(t_1)-x^\prime\| < d_0+2\epsilon$, 
    where % $0<\epsilon<\frac{1}{2}\operatorname{diam}(\mathcal{X}_{eq})$. 
    $\epsilon > 0$ sufficiently small.
 It follows that $\frac{d}{d t} W(t)<0$, $\forall   t\in [t_1,t_2)$; hence, we must have that $\lim_{t\to t_2^-} W(t)\leq W(\frac{t_2+t_1}{2})< W(t_1)$.

On the other hand, we have that $\lim_{t\to t_2^-} \sqrt{W(t)}\geq \sqrt{d_1}\|x(t_2)-h(u(t_2),w)\| \geq \sqrt{d_1} \inf_{x\prime \in \mathcal{X}_{eq} }\|x(t_2)-x^\prime\| 
=\sqrt{d_2}(d_0+\operatorname{diam}(\mathcal{X}_{eq}))$,
and 
\begin{align*}
\sqrt{W(t_1)}\leq & \sqrt{d_2}\|x(t_1)-h(u(t_1),w)\|   \\
\leq & \sqrt{d_2} \left( \inf_{x\prime \in \mathcal{X}_{eq} }\|x(t_1)-x^\prime\|+\epsilon +\operatorname{diam}(\mathcal{X}_{eq})  \right)  \\
\leq & \sqrt{d_2} (d_0+\operatorname{diam}(\mathcal{X}_{eq}) +3\epsilon). 
\end{align*}
Thus, $\lim_{t\to t_2^-} \sqrt{W(t)}+3\epsilon \sqrt{d_2}\geq \sqrt{W(t_1)}$. If we let $\epsilon\to 0$, one would have that $\lim_{t\to t_2^-} \sqrt{W(t)}\geq \sqrt{W(t_1)}$, which is a contradiction.

\vspace{.1cm}

\emph{(d) Proof of Theorem \ref{thm:stability}}. In order to demonstrate the local exponential stability of the interconnected system, we first establish an intermediate result pertaining to the stability of the ``open-loop controller'' $\dot u=F_\beta(u)$ at $u^*$. 

We note that $F_\beta(u)$ is well-defined for any $u\in\mathcal{U}_c$ since the constraints in $F_\beta(h(u,w),u)$ are always feasible by Assumption \ref{as:feasibility}. By Assumption~\ref{as:openloop}, and using 
 \cite[Lemma 5.11]{allibhoy2023control} and \cite[Theorem 5.6(iii)]{allibhoy2023control}, we deduce $F_\beta$ is differentiable at $u^*$ and its Jacobian $E=\frac{\partial F_\beta (u)}{\partial u}\mid_{u=u^*}$ is negative definite;  let  $P:=\kappa\int_0^\infty (\exp(E\zeta)^\top  \exp(E\zeta) d\zeta$ for some $\kappa>0$. 
By \cite[Theorem 4.10]{khalil2002nonlinear}, it holds that $ PE+E^\top P=-\kappa\mathbf{I}_n   $, and $
   \frac{\kappa}{2e_2} \|u-u^*\|_2^2 \leq (u-u^*)^\top P (u-u^*)\leq \frac{\kappa}{2e_1} \|u-u^*\|_2^2 $.
Let $V(u):=$ $(u-u^*)^\top P (u-u^*)$; then
\begin{align*}
    & (u - u^*)^\top P F_\beta(u)+ F_\beta(u)^\top P (u - u^*)  \\
    =& (u - u^*)^\top \left( P E + E^\top P \right) (u - u^*) \\
    &  + 2 (u - u^*)^\top P \hat{g}(u)\\
    %&= -(u - u^*)^\top Q (u - u^*) + 2 (u - u^*)^\top P \hat{g}(u)\\
    \leq & -\kappa \|u - u^*\|^2 + 2 \|u-u^*\| \|P\| \|\hat{g}(u)\| \\
    \leq & -\kappa \|u - u^*\|^2 + \frac{\kappa}{e_1}\|u-u^*\| L \|u - u^*\|^2
\end{align*}
where the last inequality holds for all $\|u - u^*\| < \rho < \min\{\delta, e_1/  L \}$. For $\dot u=\eta F_\beta(x,u)$,
it follows that
   \begin{align*}   
  \frac{1}{\eta} \dot V&=\frac{1}{\eta}\dot u^\top P (u-u^*)+\frac{1}{\eta}(u-u^*)^\top P \dot u\\
   %=&2F_\beta(x,u)^\top P (u-u^*)\\
   &=2 F_\beta(u)^\top P (u-u^*) \\
   & \quad + 2  (F_\beta (x,u)-F_\beta(u))^\top P (u-u^*)\\
& \leq  \left(-\kappa+ \frac{\kappa L}{e_1}\left\|u-u^*\right\|_2\right)\left\|u-u^*\right\|_2^2 \\
  & \quad +              2\|u-u^* \|\|P\| \| (F_\beta(u)-F_\beta(x,u))\|\\
 & \leq - \kappa s\left\|u-u^*\right\|_2^2+              \frac{\kappa}{e_1}\|u-u^* \| \| (F_\beta(u)-F_\beta(x,u))\| 
\end{align*}
where the last inequality holds  if $\|u-u^*\|\leq \frac{e_1}{L}(1-s)$, for any $s\in(s_{\min},1]$. In the proof of Lemma \ref{forward-invariant-X}, we have shown that $\frac{d}{d t} W(\tilde x, u)\leq -d_3\|\tilde{x}\|^2+\left(d_4 \ell_{h_u}+d_5\right)\|\dot{u}\|\|\tilde{x}\|$
and $\|\dot u\| \leq \eta  \ell_{F_x} \|x-h(u,w)\|+\eta\|F_\beta(u)-F_\beta(u^*)\|$. Since $  \| F_\beta(u)-F_\beta(x,u)\|\leq  \ell_{F_x} \|x-h(u,w)\| $ and $  \| F_\beta(u)-F_\beta(u^*)\|\leq \ell_{F_u} \|u-u^*\|$, one has that, if $\|u-u^*\|\leq \frac{e_1}{L}(1-s)$, $s\in(s_{\min},1]$, then
\begin{equation}\label{bound-for-Vdot}
\begin{aligned}
    \frac{d}{d t} \frac{1}{\eta} V( u)\leq  -\kappa s \left\|u-u^*\right\|_2^2+              \frac{\kappa \ell_{F_x}}{e_1}\|u-u^* \| \|\Tilde{x}\|,       
\end{aligned}
\end{equation}
\begin{equation}\label{bound-for-Wdot}\begin{aligned}
\frac{d}{d t} W(\tilde x, u) \leq  &(-d_3+ d_4\ell_{h_u} \ell_{F_x} \eta +d_5  \ell_{F_x} \eta )\|\tilde{x}\|^2\\
&+   (d_4\ell_{h_u}+d_5) \ell_{F_u} \eta \|\tilde{x}\| \|u-u^*\| .
\end{aligned}
\end{equation}

Next, define the Lyapunov function candidate $v(\tilde x, u) :=\theta \frac{1}{\eta} W\left(\tilde x, u\right)+(1-\theta) \frac{1}{\eta} V\left(u\right)$ for \eqref{eq:interconnected-system_change_variable}. 
Using \eqref{bound-for-Vdot} and \eqref{bound-for-Wdot}, the following holds if $\|u-u^*\|\leq \frac{e_1}{L}(1-s) $: 
\begin{align*}
  \frac{d}{d t} v(\tilde x, u) & %= \theta \frac{1}{\eta}  \frac{d}{dt} W\left(\tilde x, u\right)+(1-\theta) \frac{1}{\eta}\frac{d}{dt} V\left(u\right)\\
  \leq   \frac{\theta}{\eta}  \left( (-d_3+ d_4\ell_{h_u} \ell_{F_x} \eta +d_5  \ell_{F_x} \eta  )\|\tilde{x}\|^2  \right.\\
  &\left.+   (d_4\ell_{h_u}+d_5) \ell_{F_u} \eta \|\tilde{x}\| \|u-u^*\| \right)\\
  &+(1-\theta)(-\kappa s\left\|u-u^*\right\|_2^2+              \frac{\kappa \ell_{F_x}}{e_1}\|u-u^* \| \|\Tilde{x}\|)\\
  & = -\zeta^\top M  \zeta,
\end{align*}
where $ \zeta= (\| \Tilde{x} \|, \| u-u^* \|)^\top$, and $\theta\in (0,1)$, $\eta>0$. To ensure that $v(\tilde x, u) $ is a valid Lyapunov function candidate, we need  $M$ to be positive definite. Since $M$ is symmetric, the sufficient and necessary conditions for positive definiteness are $m_{11}>0$ and $m_{11} m_{22}>m_{21}m_{12}$,
which are equivalent to $ \eta<\frac{d_3}{d_4\ell_{h_u} \ell_{F_x} +d_5 \ell_{F_x}}$ and
\begin{align*}     
  \frac{s}{\eta} >&\frac{1}{4\theta(1-\theta)\kappa d_3} \left( \theta (d_4\ell_{h_u}+d_5) \ell_{F_u}+(1-\theta) \frac{\kappa \ell_{F_x}}{e_1}     \right) ^2\\
  &+\frac{ \ell_{F_x}s}{d_3}(d_4\ell_{h_u}+d_5)\\
  =&\frac{1}{4 \kappa d_3} \left( \frac{\theta}{1-\theta} \alpha_1^2+\frac{1-\theta}{\theta} \alpha_2^2+2\alpha_1\alpha_2     \right)\\
  &+\frac{ \ell_{F_x}s}{d_3}(d_4\ell_{h_u}+d_5) \geq \frac{1}{ \kappa d_3}\alpha_1\alpha_2+\frac{ \ell_{F_x}s}{d_3}
 \end{align*}
where $\alpha_1=(d_4\ell_{h_u}+d_5)\ell_{F_u}$, $\alpha_2=\frac{\kappa \ell_{F_x}}{e_1}$ for brevity. The last inequality comes from the arithmetic-geometric mean inequality~\cite{steele2004cauchy}, and equality can be attained if and only if $ \frac{\theta}{1-\theta} \alpha_1^2=\frac{1-\theta}{\theta} \alpha_2^2$, which is equivalent to $\theta=\frac{\alpha_2}{\alpha_1+\alpha_2}$. In the rest of the proof, we fix 
$$\theta=\frac{\alpha_2}{\alpha_1+\alpha_2}=\frac{\kappa \ell_{F_x}}{e_1\ell_{F_u}(d_4\ell_{h_u}+d_5)+\kappa \ell_{F_x}}.$$
Therefore, one has that  
 %\begin{align*}
$\eta< %&\frac{s d_3}{\frac{(d_4\ell_{h_u}+d_5)\ell_{F_u} \ell_{F_x}}{e_1} +({d_4}\ell_{h_u}+d_5) \ell_{F_x}s}\\ =
 \frac{sd_3e_1}{ \ell_{F_x}(d_4\ell_{h_u}+d_5)(\ell_{F_u}+e_1s)}. $    % \end{align*} 

Since we need $x(t)$ staying in $\mathcal{X}_0$, the valid range of $\eta$ is $ \eta \leq\min\left\{\frac{sd_3e_1}{ \ell_{F_x}(d_4\ell_{h_u}+d_5)(\ell_{F_u}+e_1s)} ,\alpha_0,   \frac{d_3}{2 (d_4\ell_{h_u}+d_5) \ell_{F_x} }  \right\}$, by combining the limitation for $\eta$ in Lemma \ref{forward-invariant-X}.

To conclude, we first note that
\begin{align*}
&\|\zeta\|^2=\|\tilde{x}\|^2+\|u-u^*\|^2\leq \frac{1}{d_1}W+\frac{2e_2}{\kappa}V\\
=&\frac{\eta}{\theta d_1}\frac{\theta}{\eta}W+\frac{2e_2\eta}{\kappa(1-\theta)} \frac{1-\theta}{\eta}V\leq \underbrace{\max \left\{ \frac{\eta}{\theta d_1},\frac{2e_2\eta}{\kappa(1-\theta)} \right\}}_{=r_1}  v,\\  
&\|\zeta\|^2=\|\tilde{x}\|^2+\|u-u^*\|^2\geq \frac{1}{d_2}W+\frac{2e_1}{\kappa}V\\
=&\frac{\eta}{\theta d_2}\frac{\theta}{\eta}W+\frac{2e_1\eta}{\kappa(1-\theta)} \frac{1-\theta}{\eta}V\geq \underbrace{\min \left\{ \frac{\eta}{\theta d_2},\frac{2e_1\eta}{\kappa(1-\theta)} \right\}}_{=r_2}  v.
\end{align*}
Let $\lambda_M$ be the minimum eigenvalue of $M$; then, $\dot{v}\leq-\lambda_M\|\zeta\|^2=-\lambda_M(\|\tilde{x}\|^2+\|u-u^*\|^2)\leq -\lambda_M r_2 v$. By the  
Comparison Lemma \cite[Lemma 3.4]{khalil2002nonlinear}, it follows that $v(t)\leq v(t_0)\exp(-\lambda_M r_2(t-t_0))$
if $\|u(t_0)-u^*\|\leq \frac{e_1}{L}(1-s)$, for all  $s\in(s_{\min},1]$.  Besides, we note that $\|\zeta(t)\|$ and $\|\tilde z(t)\|$ satisfy the following: 
\begin{align*}
    \|\tilde z(t)\|^2= &\|x-h(u^*,w)\|^2+\|u-u^*\|^2\\
    \leq & (\|\tilde x\|+\|h(u,w)-h(u^*,w)\|)^2+\|u-u^*\|^2\\
    \leq & \|\zeta(t)\|^2+2\ell_{h_u}\|\tilde x\|\|u-u^* \|+\ell_{h_u}^2\|u-u^* \|^2\\
    \leq & (1+\ell_{h_u}^2) \|\zeta(t)\|^2 +\ell_{h_u}(\|\tilde x\|^2+\|u-u^*\|^2)\\
    =& (1+\ell_{h_u}^2+\ell_{h_u}) \|\zeta(t)\|^2.
\end{align*}
Similarly, one can show that $\|\zeta(t)\|^2\leq (1+\ell_{h_u}^2+\ell_{h_u})    \|\tilde z(t)\|^2.$
  Hence, $\|\tilde z(t)\|\leq \sqrt{(1+\ell_{h_u}^2+\ell_{h_u})r_1v(t)}    \leq  \sqrt{(1+\ell_{h_u}^2+\ell_{h_u})r_1  v(t_0)} \exp\left(-\frac{1}{2}\lambda_M r_2(t-t_0)\right) \leq (1+\ell_{h_u}^2+\ell_{h_u})\sqrt{\frac{r_1}{r_2}} \|\tilde z(t_0)\| \exp\left(-\frac{1}{2}\lambda_M r_2(t-t_0)\right)$, for all  $t\geq t_0$, if $\|u(t_0)$ $-u^*\|\leq \frac{e_1}{L}(1-s)$, for all  $s\in(s_{\min},1]$.

\section{Representative numerical results}

In this section, we apply \eqref{eq:interconnected-system} to the problem of regulating the position of a unicycle robot to the solution of a constrained optimization problem. Specifically, 
the unicycle dynamics~\cite{lavalle2006planning}
read as $\dot{a}=v_1 \cos \theta$, $ \dot{b}=v_1 \sin \theta$, $ \dot{\theta}=v_2$, 
where $v_1, v_2 \in \mathbb{R}$ are the low-level inputs, the state is $x=(a,b,\theta)$, with $(a,b)$ the position  in a 2-dimensional plane, and $\theta \in (-\pi, \pi]$  its orientation with respect to the $a$-axis. We consider the cost functions
$\phi(u):=0.05\|u\|_2^2$, $\psi(x):=\|(a,b) -x_{\text{target}}\|_2^2$ and constraints $u\in\mathcal{U}_c:=\{(u_a, u_b)\mid -10\leq u_a,u_b\leq 10 \}$ and $\ell(x):=\|(a,b)\|_2^2-0.9\leq 0$, where 
$x_{\text{target}} = (0.6,0.8)$ is the targeted final position of the robot.
Here, we follow~\cite{cothren2022online} and focus on the error variables $\xi:=\|u-(a,b)^\top\|$ and $\bar \theta:=\operatorname{atan} 2\left(\frac{u_b-b}{u_a-a}\right)-\theta$, whose dynamics yield a globally exponentially stable equilibrium point $(\xi,\bar \theta)=(0,0)$ with the choice $v_1=k \xi \cos (\bar\theta)$ and $ v_2=k(\cos (\bar\theta)+1) \sin (\bar\theta)+k \bar\theta$, where $u=(u_a,u_b)$ is the
high-level control input given by the optimization problem. Finally, we model measurement errors in the state $x$ by instead considering $(\hat a, \hat b):= (a,b) + w$, where $w \in \mathbb{R}^2$ is a constant disturbance. %this reflects cases where the sensors are not calibrated correctly.

\begin{figure}[h!]
  \centering 
  \begin{subfigure}[Trajectory of $x(t)$ and $h(u(t),w)$]{\includegraphics[width=0.46\textwidth]{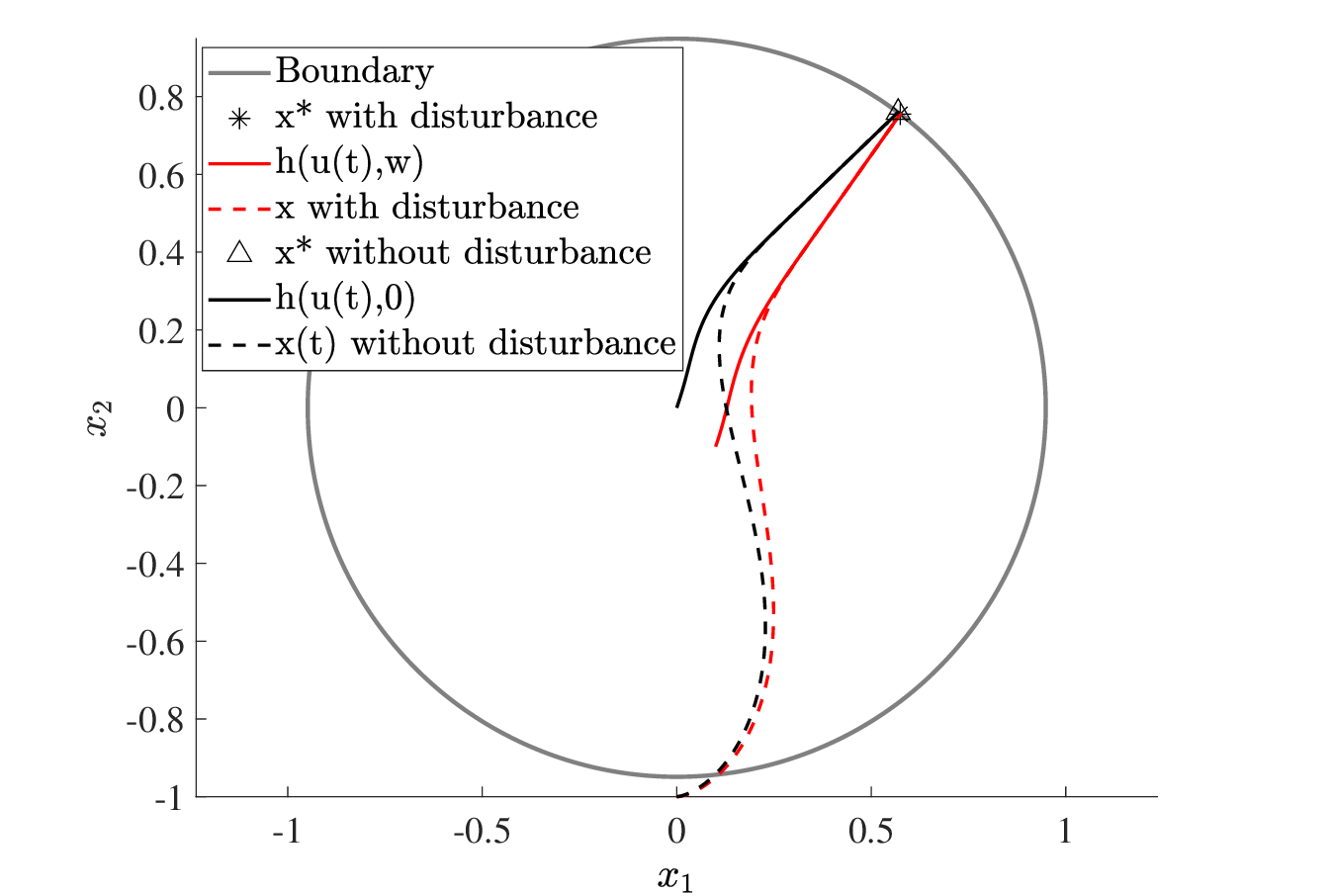}} \end{subfigure}
  \begin{subfigure}[Error plot]{\includegraphics[width=0.44\textwidth]{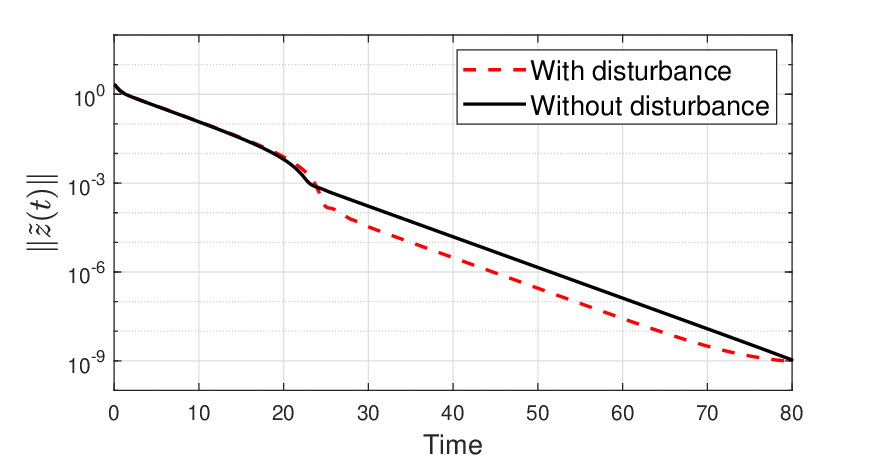}}
  \end{subfigure}
  \vspace{-.3cm}
  \caption{Trajectories for the interconnected system and error plot.}
  \label{fig:unicycle}
  \vspace{-.7cm}
\end{figure}

In Figure \ref{fig:unicycle}, we apply \eqref{eq:interconnected-system} to regulate $u$ and $x$ toward the minimizer with initial condition $x(0)= (0,-1,0)$, $u(0)=(0,0)$ and  parameters $\beta=10$, $k=2$, $\eta=0.1$. In Figure \ref{fig:unicycle}(a), we plot the trajectories of $(a,b)$ and steady-state mapping $h(u(t),w)$ with and without disturbance $w$. 
For both cases, all trajectories converge to the corresponding equilibrium $h(u^*,w)$. In Figure \ref{fig:unicycle}(b), we plot the error $\|\tilde z(t)\|$. For both cases, the error curves can be upper bounded by a plot of an exponentially decreasing function, in consistency with Theorem~\ref{thm:stability}.

\vspace{-.4cm}

\section{Conclusions}
We proposed a state feedback controller based on a continuous approximation of the projected gradient flow to regulate a dynamical system to optimal solutions of a constrained optimization problem. In particular, the optimization problem can have nonlinear inequality constraints on the system state. We  derived sufficient conditions to ensure that isolated locally optimal solutions for which the strict complementarity condition and the LICQ hold are locally exponentially stable for the closed-loop system. Future research efforts will look at extensions of our results to time-varying disturbances and to sample-data implementations of our controller.

\bibliographystyle{IEEEtran}
\bibliography{references.bib}

\end{document}